\documentclass[letterpaper]{article} 
\usepackage{amssymb}
\usepackage{amsmath}
\usepackage{booktabs} 
\usepackage{aaai24}  
\usepackage{times}  
\usepackage{helvet}  
\usepackage{courier}  
\usepackage[hyphens]{url}  
\usepackage{graphicx} 
\usepackage{natbib}  
\usepackage{caption} 
\usepackage{algorithm}
\usepackage{algorithmic}
\usepackage{color}
\usepackage{bm}
\definecolor{blue}{RGB}{0,112,192}
\definecolor{green}{RGB}{146,208,80}
\definecolor{gray}{RGB}{127,127,127}
\definecolor{vis-green}{RGB}{61,145,64}
\definecolor{red}{RGB}{255,0,0}
\definecolor{yellow}{RGB}{255,192,0}
\usepackage{subfig}

\usepackage{newfloat}
\usepackage{listings}
\DeclareCaptionStyle{ruled}{labelfont=normalfont,labelsep=colon,strut=off} 
\floatstyle{ruled}
\newfloat{listing}{tb}{lst}{}
\floatname{listing}{Listing}
\title{AMD: Autoregressive Motion Diffusion}
\author{
    Bo Han\textsuperscript{\rm 1},
    Hao Peng\textsuperscript{\rm 2},
    Minjing Dong\textsuperscript{\rm 3},
    Yi Ren\textsuperscript{\rm 1},
    Yixuan Shen\textsuperscript{\rm 4},
    Chang Xu\textsuperscript{\rm 3}\thanks{Corresponding author}
}
\affiliations{
    \textsuperscript{\rm 1}College of Computer Science and Technology, Zhejiang Univerisity,
    \textsuperscript{\rm 2}Unity China\\
    \textsuperscript{\rm 3}University of Sydney,
    \textsuperscript{\rm 4}National University of Singapore\\borishan815@gmail.com, caspian.peng@unity.cn, mdon0736@uni.sydney.edu.au\\rayeren613@gmail.com, yshe0148@gmail.com, c.xu@sydney.edu.au

}

\begin{document}

\maketitle

\begin{abstract}
Human motion generation aims to produce plausible human motion sequences according to various conditional inputs, such as text or audio. Despite the feasibility of existing methods in generating motion based on short prompts and simple motion patterns, they encounter difficulties when dealing with long prompts or complex motions.
The challenges are two-fold: 1) the scarcity of human motion-captured data for long prompts and complex motions. 2) the high diversity of human motions in the temporal domain and the substantial divergence of distributions from conditional modalities, leading to a many-to-many mapping problem when generating motion with complex and long texts. 
In this work, we address these gaps by 1) elaborating the first dataset pairing long textual descriptions and 3D complex motions (HumanLong3D), and 2) proposing an autoregressive motion diffusion model (AMD). Specifically, AMD integrates the text prompt at the current timestep with the text prompt and action sequences at the previous timestep as conditional information to predict the current action sequences in an autoregressive iterative manner.
Furthermore, we present its generalization for X-to-Motion with “No Modality Left Behind”, enabling the generation of high-definition and high-fidelity human motions based on user-defined modality input.
\end{abstract}
\section{Introduction}
Human motion generation is a crucial task in computer animation and has applications in various fields including gaming, robots, and film. Traditionally, new motion is accessed through motion capture in the gaming industry, which can be costly. As a result, automatically generating motion from textual descriptions or audio signals can be more time-efficient and cost-effective. 
Related research work is currently flourishing, exploring human motion generation from different modalities \cite{tevet2022human, zhang2022motiondiffuse, tseng2022edge, li2021ai}. 

Current text-based conditional human motion synthesis approaches have demonstrated plausible mapping from text to motion \cite{petrovich2022temos, tevet2022human, zhang2022motiondiffuse, guo2022generating, zhang2023t2m}. They are mainly divided into three categories: 

Latent space strategy \cite{petrovich2022temos,ahuja2019language2pose,tevet2022motionclip}: This is typically done by separately learning a motion Variational AutoEncoder (VAE) \cite{kingma2013auto} and a text encoder, and then constraining them to a compatible latent space using the Kullback-Leibler (KL) divergence loss. However, since the distributions of natural language and human motion are vastly different, forcibly aligning these two simple Gaussian distributions can result in misalignments and diminished generative diversity.

Diffusion-based approach \cite{tevet2022human,zhang2022motiondiffuse,chen2023mld}: diffusion models \cite{ho2022classifier,song2020score} have recently attracted significant attention and have shown remarkable breakthroughs in various areas such as video \cite{luo2023videofusion}, image \cite{ramesh2022hierarchical}, and 3D point cloud generation \cite{han2023zero3d}, etc. Current motion generation methods based on diffusion models \cite{tevet2022human, zhang2022motiondiffuse, chen2023mld} have achieved exceptional results using different denoising strategies.

Typically, MDM \cite{tevet2022human} proposes a motion diffusion model on raw motion data to learn the relationship between motion and text conditions. However, these models tend to only generate single motions or contain several motion sequences and are often inefficient for complex long texts.
Autoregressive method \cite{gopalakrishnan2019neural, pavllo2020modeling, TEACH:3DV:2022}: they can process varying motion lengths, tackling the issue of fixed motion duration. However, their single-step generation methods often rely on traditional VAE models \cite{kingma2013auto}, which are less effective than diffusion models.
Despite the progress made by existing methods, text-based conditional human motion generation remains a challenging task for several reasons:
\begin{itemize}
    \item Lack of enough motion-captured data: At present, there are few widely used text-to-motion datasets \cite{plappert2016kit, guo2022generating, punnakkal2021babel}, which mostly contain simple motions and are deficient in long text prompts, i,e., "he is flying kick with his left leg". 
    \item Weak correlation: Due to the differing distributions of natural language and human motion, resulting in a multiple mapping problem \cite{tevet2022human}. This issue is further exacerbated when generating long text-based human motions.
\end{itemize}
To address the aforementioned limitations and challenges, we propose Autoregressive Motion Diffusion model (AMD) that can generate motion sequences with complex long content, variable duration, and multiple modalities. It leverages the generative capabilities of the diffusion model and the temporal modeling strengths of the autoregressive model.
Considering the high dimensionality of complex long motion sequences, in order to better capture the dependencies between texts and motions in long sequences, AMD combines the text description at the current timestep with the text description and motion information at the previous timestep as conditional information to predict the motion sequence at the current timestep. 
AMD continuously employs the diffusion method to synthesize the corresponding motion sequence from the previous timestep and finally can generate the motion sequences of all texts. Besides, we explicitly design several geometric constraints to encourage matching physical realism in including height, joint position, joint rotation, joint velocity, and sliding loss.
To address the scarcity of human motion-captured data for long prompts and complex motions, we have developed HumanLong3D - the first dataset to pair long textual descriptions with complex 3D human motions, i.e., "A person is doing martial art action raising knees and stretching feet, and then the person performs step forward with his right foot". The dataset comprises 158,179 textual descriptions and 43,696 3D human motions. It encompasses a broad spectrum of complex motion types. Importantly, it features annotations for motion coherence.
In addition, we have also developed the HumanMusic dataset to evaluate the generation effect across different modalities. This dataset pairs 137,136 motions with corresponding audio data. They all follow the format of the HumanML3D dataset \cite{guo2022generating}. The
codes for AMD and demos can be found in the Supplementary Materials.

In summary, our contributions include:
\begin{itemize}
    \item We propose a novel continuous autoregressive diffusion model that combines state-of-the-art performance for generating complex and variable motions on long texts. 
    \item We construct two large-scale cross-modal 3D human motion datasets HumanLong3D and HumanMusic, which could serve as the benchmarks for future cross-modal motion generation.
    \item Our proposed AMD achieves impressive performances on the HumanML3D, HumanLong3D, AIST++, and HumanMusic datasets, which highlights its ability to generate high-fidelity motion given different modality inputs.
\end{itemize}
\section{Background}
Human motion generation has been an active area of research for many years \cite{badler1993simulating}. Early work in this field focused on unconditional motion generation \cite{rose1998verbs, mukai2005geostatistical, ikemoto2009generalizing}, with some studies attempting to predict future motion based on an initial pose or starting motion sequence \cite{o1980model, gavrila1999visual}. Statistical models such as Principal Component Analysis (PCA) \cite{ormoneit2005representing} and Motion Graphs \cite{min2012motion} were commonly used for these generative tasks. 
The development of deep learning has led to the emergence of an increasing number of sophisticated generative architectures \cite{kingma2013auto, vaswani2017attention, goodfellow2020generative, kingma2018glow, ho2020denoising}. These advanced generative models have encouraged researchers to explore conditional motion generation.
Conditional human motion generation can be modulated by a variety of signals that describe the motion, with high-level guidance provided through various means such as action classes \cite{petrovich2022temos}, audio \cite{aristidou2022rhythm}, and natural language \cite{ahuja2019language2pose, petrovich2022temos}.
\subsection{Text-To-Motion}
Due to the language descriptors are the most user-friendly and convenient. Text-to-motion has been driving and dominating research frontiers.
In recent years, the leading approach for the Text-to-Motion task is to learn a shared latent space for language and motion. 
JL2P \cite{ahuja2019language2pose} learns from the KIT-ML dataset \cite{plappert2016kit} with an auto-encoder, limiting one-to-one mapping from text to motion. 
TEMOS \cite{ahuja2019language2pose} and T2M \cite{guo2022generating} propose using a VAE \cite{kingma2013auto} to map a text prompt into a normal distribution in latent space. Recently, MotionCLIP \cite{tevet2022motionclip} has leveraged the shared text-image latent space learned by CLIP to expand text-to-motion beyond data limitations and enable latent space editing. However, due to the inconsistency of the two data distributions of natural language and human motion, it is very difficult to align them in the shared latent space. 
Diffusion Generative Models \cite{sohl2015deep} achieve significant success in the image synthesis domain, such as Imagen \cite{saharia2022photorealistic}, DALL2 \cite{ramesh2022hierarchical} and Stable Diffusion \cite{rombach2022high}. 
Inspired by their works, most recent methods \cite{tevet2022human, zhang2022motiondiffuse,chen2023mld} leverage diffusion models for human motion synthesis. 
MotionDiffuse \cite{zhang2022motiondiffuse} is the first work to generate human motion that corresponds to text utilizing a diffusion model. Recently, MDM \cite{tevet2022human} has been proposed, which operates on raw motion data to learn the relationship between motion and input conditions. Inspired by Stable Diffusion \cite{rombach2022high}, MLD \cite{chen2023mld} implements the human motion diffusion process in the latent space. Despite their ability to produce exceptional results, these models are typically limited to short text descriptions and simple motions.
Additionally, several works \cite{gopalakrishnan2019neural, pavllo2020modeling, TEACH:3DV:2022} have been developed based on the concept of autoregression, which can generate human actions of any length. ARDMs~\cite{HoogeboomGBPBS22} combines the order-agnostic autoregressive model and the discrete diffusion model, which eliminates the need for causal masking of model representations and enables fast training, allowing it to scale favorably to high-dimensional data.
Consequently, for long text prompts, we combine the advantages of the diffusion model in generating motion for short text descriptions with the concept of autoregression to achieve superior human motion results for continuous long text.
\subsection{Motion Datasets}
Common forms of description for human motion data are 2D keypoints, 3D keypoints, and statistical model parameters \cite{yu2020humbi, cai2022humman}. 
For the text-conditioned motion generation task, KIT \cite{plappert2016kit} is the first 3D human motion dataset with matching text annotations for each motion sequence. HumanML3D \cite{guo2022generating}  provides more textual annotation for some motions of AMASS \cite{mahmood2019amass}. They are also our focus in the text-to-motion task. Babel \cite{punnakkal2021babel} also collects motions from AMASS \cite{mahmood2019amass} and provides action and behavior annotations, it annotates each frame of the action sequence, thereby dividing compound actions into simple action groups. In this paper, we use the HumanML3D dataset to evaluate the proposed methods for simple motions and short prompts. In addition, we collected and labeled pairs of complex motion data and text prompts (HumanLong3D). More importantly, we provided temporal motion-coherence information to support long text-to-motion generation tasks.
\subsection{Audio-To-Motion}
Generating natural and realistic human motion from audio is also a challenging problem. Many early approaches follow a motion retrieval paradigm \cite{fan2011example, lee2013music}. A traditional approach to motion synthesis involves constructing motion maps. New motions are synthesized by combining different motion segments and optimizing transition costs along graph paths \cite{safonova2007construction}. 
More recent approaches employ RNN \cite{tang2018dance, alemi2017groovenet, huang2020dance}, GANs \cite{lee2019dancing, sun2020deepdance}, Transformer \cite{li2022danceformer, li2021ai, siyao2022bailando}, and CNN \cite{holden2016deep} models to map the given music to a joint sequence of the continuous human pose space directly. Such methods would regress to nonstandard poses that are beyond the dancing subspace during inference. In contrast, our proposed method does not produce the phenomenon of limb drift.
\section{Our Approach}
In this section, we first introduce the problem formulation for long text-to-motion. To enable adaptive motion generation for different text descriptions, we propose the inclusion of a motion duration prediction network to approximate the duration. To generate human motions that correspond to continuous long text descriptions, we establish an autoregressive motion diffusion model.
\subsection{Problem Description}
To generate complex motion sequences with long-term text prompts, we propose to feed multiple text prompts in order. Given $N$ text prompts $S^{1:N} = \left\{S^1, S^2,\ldots, S^N \right\}$, the model is required to generate $N$ motion segments $X^{1:N} = \left\{X^1, X^2,\ldots, X^N \right\}$ consistent with the text descriptions, where $N$ denotes the number of motion segments involved in the entire motion sequence.
Each motion segment is defined as $X^i = \left\{x^1,x^2,\ldots,x^{F^i} \right\}$, where $F^i$ is the total number of frames of the motion segment $X^i$ and $x^j$ denotes the 3D human body pose representation of the $j$-th frame. 
It is imperative that each generated motion segment and the corresponding number of motion frames adhere to the specifications outlined in the text prompt. Additionally, a seamless transition from $X^{i-1}$ to $X^i$ is crucial for the generation of high-fidelity motion.
\subsection{Overall Framework}
\begin{figure*}[htb]
    \centering
    \includegraphics{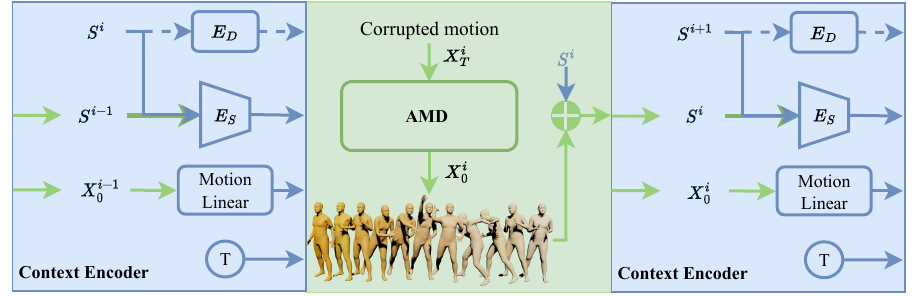}
    \caption{Overview of the Autoregressive Motion Diffusion model. Given the current text prompt $S^i$, the last text prompt $S^{i-1}$, and motion $X_0^{i-1}$ \textcolor{green}{(green arrow)}, we first encode the context information \textcolor{blue}{(blue block)}. Then, we feed the input conditions and corrupted motion $X_T^i$ to AMD Mudule (Fig. \ref{fig: MDM}) to generate the original cleaned motion $X_0^i$. Afterward, we send the current text prompt $S^{i}$ and motion $X_0^i$ to the next time step. Iteratively, we can obtain motion sequences for long text prompts.}
    \label{fig: overall}
\end{figure*}

It is important to note that daily human motions encompass not only simple, single motions but also complex, prolonged motions that more accurately reflect real-life scenarios.
Specifically, given a series of semantic prompts $S^{1:N}$, a series of randomly sampled temporal motion sequences $X_T^{1:N}\sim\mathcal{N}(0, I)$ obeying the standard normal distribution, and a maximum noise scale $T\in\mathbb{N}$ where each semantic prompt $S^i$ describes a single and distinct motion. 
Our goal is to generate noise-free temporal motion sequences $X_0^{1:N}$, which are guided by the semantic prompts, with smooth transitions between adjacent motions $X_0^{i-1}$ and $X_0^i$.
The overall process is illustrated in Fig \ref{fig: overall}, and each pair of blue and green blocks represents each step of the AMD model. $S^{1:N}$ employs the model iteratively to synthesize motion $X_0^{1:N}$. The blue block represents the context encoder and the green block is the motion diffusion module.
\subsubsection{Motion Duration Prediction Network}
\label{sector: motion duration}
Given a semantic prompt $S^i$, the duration of $X^i$ may vary. For instance, in the HumanML3D dataset, the prompt "a man kicks something or someone with his left leg" corresponds to 116 frames, while the prompt "a person squats down then jumps" corresponds to 35 frames. Consequently, we propose predicting the motion duration in order to generate motions with adaptive length. 
Similar to T2M, we use probability density estimation to determine the number of frames required for the motion synthesis based on text prompts. 
Due to the diversity of the motion duration, it is more reasonable to model the mapping problem as a density estimation problem than directly regressing the specific value. 
By utilizing the semantic prompt $S^i$ as input for the motion duration prediction network, a probability density estimation is conducted on the discrete group encompassing all possible motion durations $L = \left\{ L_{min}, L_{min}+1,\ldots, L_{max}\right\}$. The loss function of the network is designed as the cross-entropy loss of multi-classification.
\subsubsection{Context Encoder}
It includes the motion duration prediction network $E_D$, the semantic conditional encoder $E_S$, and the motion linear layer. 
The CLIP model \cite{radford2021learning} is utilized as the semantic conditional encoder. Given that our primary focus is on long text-to-motion generation, it is necessary to consider timing-related information associated with long texts. To this end, we encode the previous motion $X_0^{i-1}$ by the motion linear layer to obtain $z_{m}^{i-1}$ and encode semantic information $S^{i-1}$ by the semantic conditional encoder $E_S$ to obtain $z_{c}^{i-1}$. These are then concatenated to form the final prior condition feature $z_{past}^{i-1}$. 
Simultaneously, the current semantic information is input into the motion duration prediction network and semantic conditional encoder to obtain $F^i$ and $z_c^i$, respectively.
In order to avoid overfitting, we perform a random mask on the semantic conditional information $z_c^i$. For the corrupted motion $X_t^i$, the same motion linear layer is utilized to obtain the encoded information $z_m^i$. 
We feed the diffusion time scale $t$ to a Multi-layer Perceptron (MLP) to obtain the time embedding $z_t$. The final condition information $z$ is defined as follows:
\begin{equation}
    z = C(C(C(z_{m}^{i-1}, z_{c}^{i-1})+RM(z_c^i), z_t), z_m^i, PE(F^i))
    \label{equ: condition embedding}
\end{equation}
where $C$ represents the concatenation operation, $RM$ denotes random mask, and position embedding refers to potion embedding. It is important to note that during training, we utilize the actual motion duration present in the dataset, whereas, during the inference phase, the predicted duration information is used.
\subsubsection{AMD Module}
\begin{figure*}[htb]
    \centering
    \includegraphics{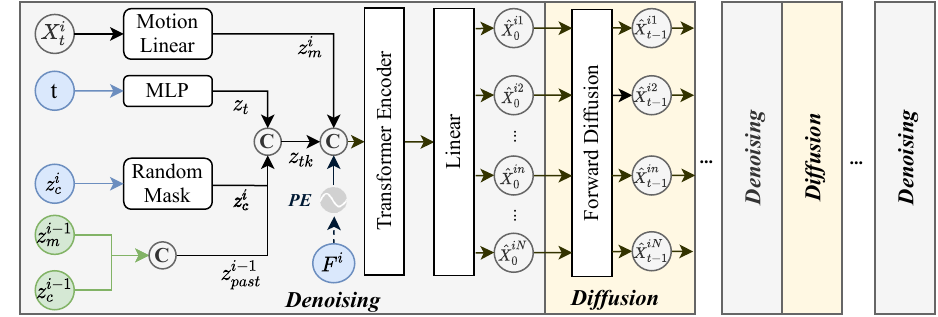}
    \caption{AMD Module. The \textcolor{gray}{gray blocks} denote the denoising process, while the \textcolor{yellow}{yellow blocks} represent the diffusion process. Within the AMD module, they appear in pairs T times (with the exception of the last one).}
    \label{fig: MDM}
\end{figure*}
The network architecture of the AMD module is depicted in Fig \ref{fig: MDM}. The denoising process (gray) and the diffusion process (yellow) span a total of $T$ timesteps, where $T$ represents the pre-defined maximum time scale. We directly predict the original cleaning motion clips during the denoising process, while the diffusion process operates in the opposite direction. The single step of the diffusion process is essentially the transfer process from $X_{t-1}^i$ to $X_{t}^i$, as defined in the following:
\begin{equation}
    \label{equ: DiffForwardSingle}
    q(X_t^i|X_{t-1}^i)=\mathcal{N}(X_t^i;\sqrt{1-\beta_t}X_{t-1}^i,\beta_t I),
\end{equation}
Where $\beta_t$ is pre-defined to regulate the magnitude of noise addition. The transition probability, denoted as Equation~\ref{equ: DiffForwardTotal}, from $X_{0}^i$ to $X_{t}^i$ can be derived using Equation~\ref{equ: DiffForwardSingle} in conjunction with the multiplication formula for Gaussian distribution, where $\alpha_t=1-\beta_t$ and $\bar{\alpha}_t=\prod_{s=1}^{t}\alpha_s$.
\begin{equation}
    \label{equ: DiffForwardTotal}
    q(X_{t}^i|X_0^i)=\prod_{t=1}^{t}q(X_t^i|X_{t-1}^i)=\mathcal{N}(X_t^i;\sqrt{\bar{\alpha}_t}\hat{X}_0^i,(1-\bar{\alpha}_t)I).
\end{equation}
The single step of the denoising process is essentially the transfer process from $X_{t}^i$ to $X_{t-1}^i$, the transfer strategy requires a network with parameters $\theta$ to learn the sampling distribution as:
\begin{equation}
\begin{aligned}
    \label{equ: DiffBackwardSingle}
    p_{\theta}(X_{t-1}^i|X_t^i, z) &=q(X_{t-1}^i|\hat{X}_0^{\theta}(X_t^i, t, z))\\
    & = \mathcal{N}(X_t^i;\sqrt{\bar{\alpha}_t}\hat{X}_0^{\theta}(X_t^i, t, z),(1-\bar{\alpha}_t)I),\\
\end{aligned}
\end{equation}
where $\hat{X}_0^{\theta}(X_t^i, t, z)$ represents the neural network with parameter $\theta$, which takes in $X_t^i$, $t$, and conditional information $z$ as input.

After predicting $\hat{X}_0$ through $X_t$ using the inverse diffusion network, it is necessary to perform the forward diffusion process according to the calculation method outlined in Equation~\ref{equ: DiffForwardTotal} to obtain the noise motion clips $X_{t-1}$ of the subsequent noise scale, thus completing the single-step inverse diffusion process. In summary, the algorithm randomly samples the noise motion segment $X_T$ with the largest noise scale from the standard normal distribution $\mathcal{N}(0, I)$ and iteratively executes the single-step inverse diffusion model. 
Each prediction approximately coarse cleaned action sequence $\hat{X}_0$ before forward diffusion is performed to obtain the action segment of the next noise scale. This process continues until the noise scale reaches 0 and returns to the original cleaned action sequence $X_0$.
\subsubsection{Explicit Constraints}
In the AMD Module, the original motion sequence is explicitly predicted. To enhance physical realism, we design multiple geometric constraints, including ${L}_{h}$, ${L}_{p}$, ${L}_{r}$, ${L}_{v}$, and ${L}_{f}$. The loss function of each part of human motion is defined as the L2 loss between the predicted values and the ground truth.

Among them, ${L}_{h}$ represents the height loss.
${L}_{p}$ represents the joint position loss.
${L}_{r}$ represents the joint rotation loss. 
${L}_{v}$ represents the joint velocity loss.
${L}_{f}$ represents the sliding foot loss.
Finally, the loss function is defined as:
\begin{equation}
\scalebox{0.90}{$ \displaystyle
    \label{equ:L_train}
    \mathcal{L}_{train} = \lambda_{h}\mathcal{L}_{h} + \lambda_{p}\mathcal{L}_{p} + \lambda_{r}\mathcal{L}_{r} + \lambda_{v}\mathcal{L}_{v} + \lambda_{f}\mathcal{L}_{f}
    $}
\end{equation}
where $\lambda$ denotes the coefficients to balance the loss terms.

With the proposed AMD, we are able to generate motion sequences according to ordered text prompts iteratively. Specifically, we commence from the first prompt $S^1$ and utilize the AMD to synthesize the corresponding clean motion sequence $X_0^1$. The remaining high-fidelity motion sequences $X^{2:N}_{0}$ can be synthesized using prior condition information as well as $S^{2:N}$. Ultimately, a coherent motion sequence of any length can be synthesized.
\begin{figure*}[ht]
    \centering
    \includegraphics[width=0.95\linewidth]{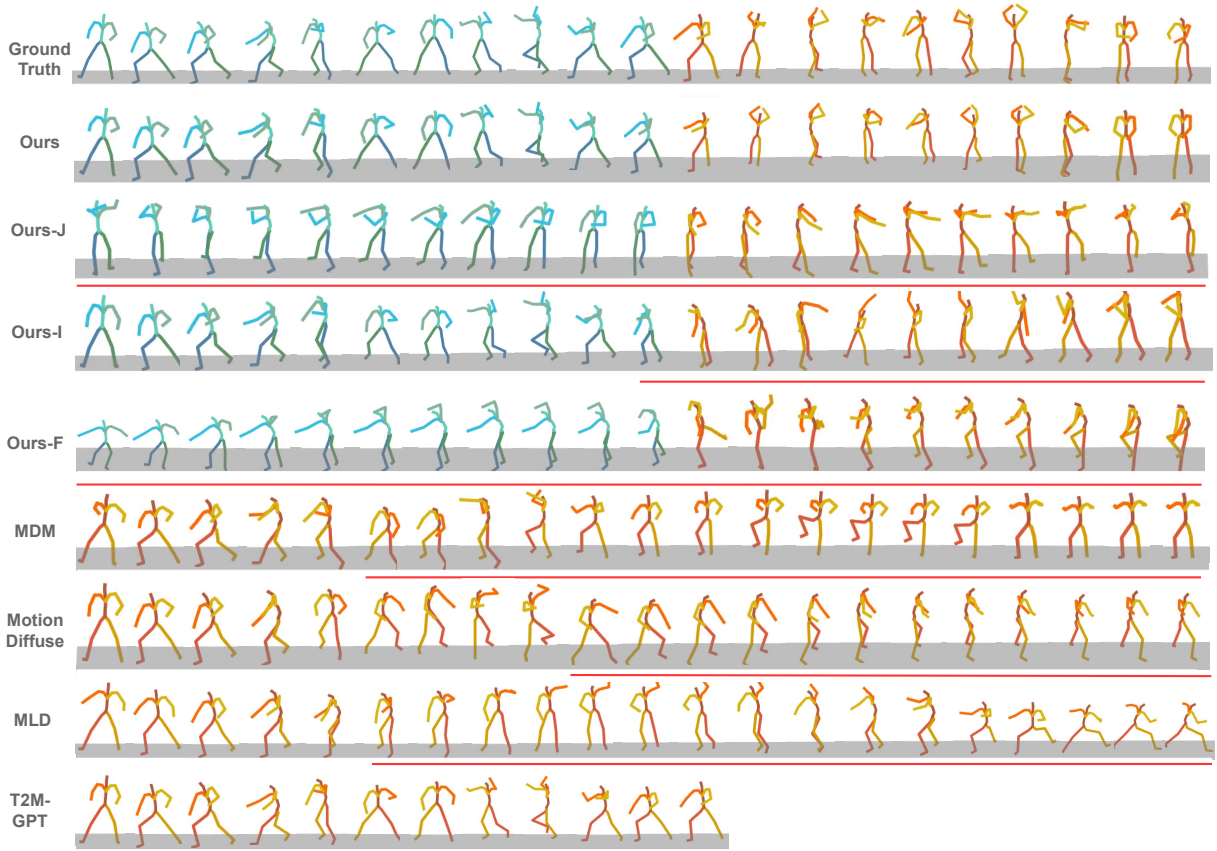}
    \caption{Result for compound motion synthesis (\textcolor{blue}{blue}: "there is a man doing left smash right cover." \textcolor{yellow}{yellow}: “then he steps forward and turn around”). The part (\textcolor{red}{red line}) indicates a discrepancy between the generated motion and the ground truth.}
    \label{fig: visualization for compound motion synthesis}
    \vspace{-0.1in}
\end{figure*}
\section{Experiments}
\subsection{Datasets and Evaluation Metrics}
\noindent\textbf{HumanLong3D} \quad
We collected motion data using motion capture equipment and online sources and annotated each motion sequence with various semantic labels to create the HumanLong3D dataset. The data format of the HumanLong3D dataset is consistent with that of HumanML3D, and it additionally includes coherence information for motion sequences to support temporal motion generation tasks. 

\noindent\textbf{HumanML3D}\quad
The dataset involves the textual re-annotation of motion capture data from the AMASS \cite{mahmood2019amass} and HumanAct12 \cite{guo2020action2motion}, comprising 14,616 motions annotated with 44,970 textual descriptions. The comparison of datasets is shown in Table~\ref{tab:dataset}.

\noindent\textbf{HumanMusic} \quad
We collected dance videos from online sources and extracted the pose parameters of the dancers in the videos, converting the motion data into the HumanML3D format. The frame
rate of each video was normalized to 20 FPS and the sampling rate of the accompanying music was
standardized to 10240Hz.
In total, we obtained 137,136 paired dance and music data samples, with each dance sample consisting of 200 frames. Music features were extracted using the publicly available
audio processing toolbox Librosa~\cite{jin2017towards}.

\noindent\textbf{AIST++} \quad
This dataset~\cite{li2021ai} comprises 992 high-quality 3D pose sequences in SMPL format \cite{loper2015smpl}, captured at 60 FPS, with 952 sequences designated for training and 40 for evaluation. We followed the approach of Bailando \cite{siyao2022bailando} to partition the dataset.
\begin{table}[hb]
\small
    \centering
    \begin{tabular}{c|c|c|c}
    \toprule
    Dataset & Motion & Textual descriptions  & Duration\\
    \midrule
    KIT-ML    &  3911 & 6248  & 10.33h \\
    HumanML3D &  14616 &44970  &28.59h \\
    HumanLong3D & 43696 &158179  &85.87h \\
    \bottomrule
    \end{tabular}
    \caption{Text-to-motion dataset description}
    \label{tab:dataset}
\end{table}

\noindent\textbf{Evaluation Metrics} \quad
For text-to-motion evaluation, we employ metrics consistent with existing methods \cite{tevet2022human, zhang2022motiondiffuse}. Specifically, (a) Frechet Inception Distance (FID) is used as the primary metric to evaluate the feature distributions between generated and real motions in feature space \cite{guo2022generating}, and (b) R-Precision (top 3) calculates the top 3 matching accuracy between text and motion in feature space. (c) \textit{MultiModal Dist} calculates the distance between motions and texts. (d) \textit{Diversity} measures variance through features. (e) \textit{MultiModality} assesses the diversity of generated motions for the same text. 
For music-to-dance evaluation, we employ metrics consistent with existing methods \cite{siyao2022bailando}. 
\subsection{Implement Details}
\begin{table*}[htb]
    \centering
    \begin{tabular}{cccccc}
    \toprule
    Method & R-Precision(Top3)$\uparrow$ & FID$\downarrow$ & MultiModal Dist$\downarrow$ & Diversity$\rightarrow$ & MultiModality$\uparrow$ \\
    \midrule
    Ours-J  & \(0.120^{\pm.006}\) & \(12.679^{\pm.063}\) & \(9.991^{\pm.047}\) & \(4.683^{\pm.117}\) & \(\textcolor{red}{3.646^{\pm.285}}\)\\
    Ours-I & \(0.142^{\pm.003}\) & \(\textcolor{vis-green}{0.827^{\pm.036}}\) & \(7.989^{\pm.056}\) & \(4.329^{\pm.065}\) & \(2.249^{\pm.206}\) \\
    Ours-F & \(0.132^{\pm.008}\) & $\textcolor{blue}{0.576^{\pm.042}}$ & \(7.937^{\pm.023}\) & \(4.312^{\pm.052}\) & \(2.257^{\pm.269}\) \\ 
    MDM \cite{tevet2022human} &\(0.096^{\pm.005}\) & \(27.348^{\pm.349}\) & \(8.203^{\pm.039}\) & \(0.781^{\pm.040}\) & \(0.547^{\pm.037}\)\\
    MotionDiffuse \cite{zhang2022motiondiffuse} &\(\textcolor{vis-green}{0.156^{\pm.002}}\) & \(6.860^{\pm.113}\) & \({8.783^{\pm.028}}\) & \(\textcolor{blue}{4.529^{\pm.076}}\) & \(2.409^{\pm.204}\)\\
    T2M-GPT \cite{zhang2023t2m} &\(\textcolor{red}{0.159^{\pm.005}}\) & \(1.249^{\pm.026}\) & \(\textcolor{red}{7.653^{\pm.019}}\) & \({4.895^{\pm.047}}\) & \(\textcolor{blue}{3.093^{\pm.104}}\)\\
    MLD \cite{chen2023mld} &\(0.144^{\pm.002}\) & \(3.843^{\pm.058}\) & \(\textcolor{vis-green}{7.847^{\pm.011}}\) & \(\textcolor{vis-green}{4.365^{\pm.033}}\) & \(\textcolor{vis-green}{2.831^{\pm.072}}\) \\
    \midrule
    Ours  & $\textcolor{blue}{0.158^{\pm.006}}$ & $\textcolor{red}{0.225^{\pm.013}}$ & $\textcolor{blue}{7.745^{\pm.029}}$ & \(\textcolor{red}{4.515^{\pm.135}}\) & \(1.242^{\pm.118}\) \\
    GT   & \(0.162^{\pm.005}\) & \(0.003^{\pm.001}\) & \(7.119^{\pm.013}\) & \(4.456^{\pm.073}\) & - \\
    \bottomrule
    \end{tabular}
    \vspace{-0.1in}
    \caption{Compound motion generation evaluation on HumanLong3D Dataset. For each metric, we repeat the evaluation 20 times (except MultiModality runs 5 times). \textcolor{red}{Red}, \textcolor{blue}{Blue}, and \textcolor{vis-green}{Green} indicate the first, the second, and the third best result.}
    \label{tab: Compound motion synthesis evaluation on HumanLong3D Dataset}
    \vspace{-0.15in}
\end{table*}

\noindent\textbf{Motion Representation} \quad
Our motion representation adopts the same format as HumanML3D, i.e., $X\in\mathbb{R}^{263\times F}$. Each frame of motion is 263-dimensional data, including the position, linear velocity, angular velocity, joint space rotation of three-dimensional human joints, and label information for judging whether the foot joints are still.
Since images are often represented as $I\in\mathbb{R}^{W\times H\times C}$.

\noindent\textbf{Motion Duration Prediction Network} \quad
$L_{min}$ is set to 10 and $L_{max}$ is 50, each unit increment corresponds to 4 motion frames, i.e., 0.2s motion duration, so the duration prediction range covers the lower bound of 2s and the upper bound of 9.8s of the data samples. The motion duration prediction network is pretrained, with the motion duration prediction network being used only during inference. 

\noindent\textbf{AMD Module} \quad
We set the maximum noise scale $T$ to be 1000, the coefficient $\beta_{1:T}$ is set to a linear increment from $10^{-4}$ to 0.02, latent vector dimensions are 512, the number of layers of the motion encoder is 6, and the number of heads of the multi-head attention mechanism is set to 6, the learning rate is fixed at $10^{-4}$, the number of training steps is 200000, and we use AdamW optimizer.

\noindent\textbf{Other Settings} \quad
The output dimension of the motion linear layer and the latent vector dimension of the AMD module are both 512. The semantic conditional encoder adopts the CLIP-ViT-B/32 checkpoint.
During inference, the semantic prompt $S^i$ is input into the motion duration prediction network $E_D$ to obtain the estimated value $F^i$ of the motion sequence duration, which is used to determine the timing dimension for motion sequence sampling.
\subsection{Comparisons on Compound Motion}
We compare compound motion generation with SOTA methods \cite{chen2023mld, tevet2022human, zhang2022motiondiffuse, zhang2023t2m}. Since the HumanML3D dataset does not contain motion coherence information, we conducted this experiment only on the HumanLong3D dataset, and we divided the dataset into training, test, and validation sets using a ratio of 0.85:0.10:0.05. Additionally, we designed three benchmarks based on TEACH \cite{TEACH:3DV:2022}: 1) Joint prediction (ours-J): The long semantic prompt $S^{i-1:i}$ formed by the combination of two coherent prompts are used as the input of the diffusion model, and a coherent time-series motion sequence $X_0^{i-1:i}$ is obtained by direct joint prediction. 2) Linear interpolation (ours-I): This method interpolates the results of two independent motion synthesis. 3) Motion filling (ours-F): Similar to linear interpolation, two independent motion synthesis are required to obtain $X_0^{i-1}$ and $X_0^i$, and the time window is set to 10\% of the motion sequence duration. All frame data except for the time window are fixed, and the frame data within the time window are filled with random normal distribution noise. The coherent motion sequence is then restored through the denoising process.

As shown in Table~\ref{tab: Compound motion synthesis evaluation on HumanLong3D Dataset},  
Among the five evaluation metrics, AMD achieved top 3 performance in four of them, with FID and Diversity, the primary metrics for motion generation quality, ranking first, demonstrating its superiority in the long text-to-compound motion generation task. Notably, AMD outperformed other methods by a significant margin in the FID metric. While the “Ours-J” scheme, despite having the highest Multimodality index, performed poorly in terms of FID, indicating its inability to generate reasonable human movements. In cases where synthesis quality is extremely low, high diversity in Multimodality may indicate that the synthesized actions are chaotic and irregular.

As illustrated in Fig~\ref{fig: visualization for compound motion synthesis}, compared to the ground truth, AMD keeps with the highest degree of similarity, while MDM, MotionDiffuse, and MLD all exhibited varying degrees of limb stiffness. Although T2M-GPT achieves results comparable to the ground truth in the first half of motion generation, its performance deteriorates in longer text-to-motion generation tasks. This is due to its premature prediction of the terminator, resulting in a lack of corresponding motion sequences for the second half of the text. For T2M-GPT, we also tried to separate the long texts into short texts and generated single motion on the short texts individually. T2M-GPT performs well in single motion generation tasks but struggles with compound motion generation tasks. Additionally, when generating long text-to-compound motion, dividing the long text and generating it separately often results in unnatural transitional motion clips.

\begin{figure}[htb]
\centering
\subfloat[the person picks an object up off the floor with their left hand]{
\label{fig:subfig:a}
\includegraphics[width=0.95\linewidth]{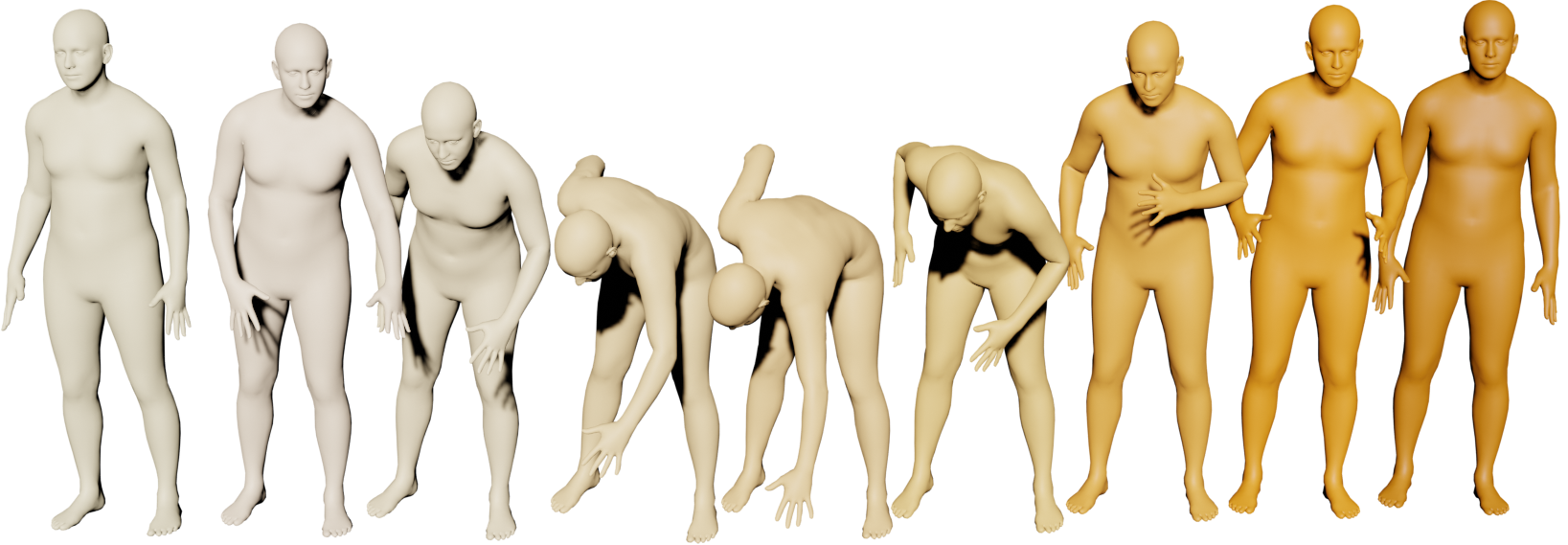}
}

\subfloat[a person throws an object with his right hand.]{
\label{fig:subfig:a}
\includegraphics[width=0.95\linewidth]{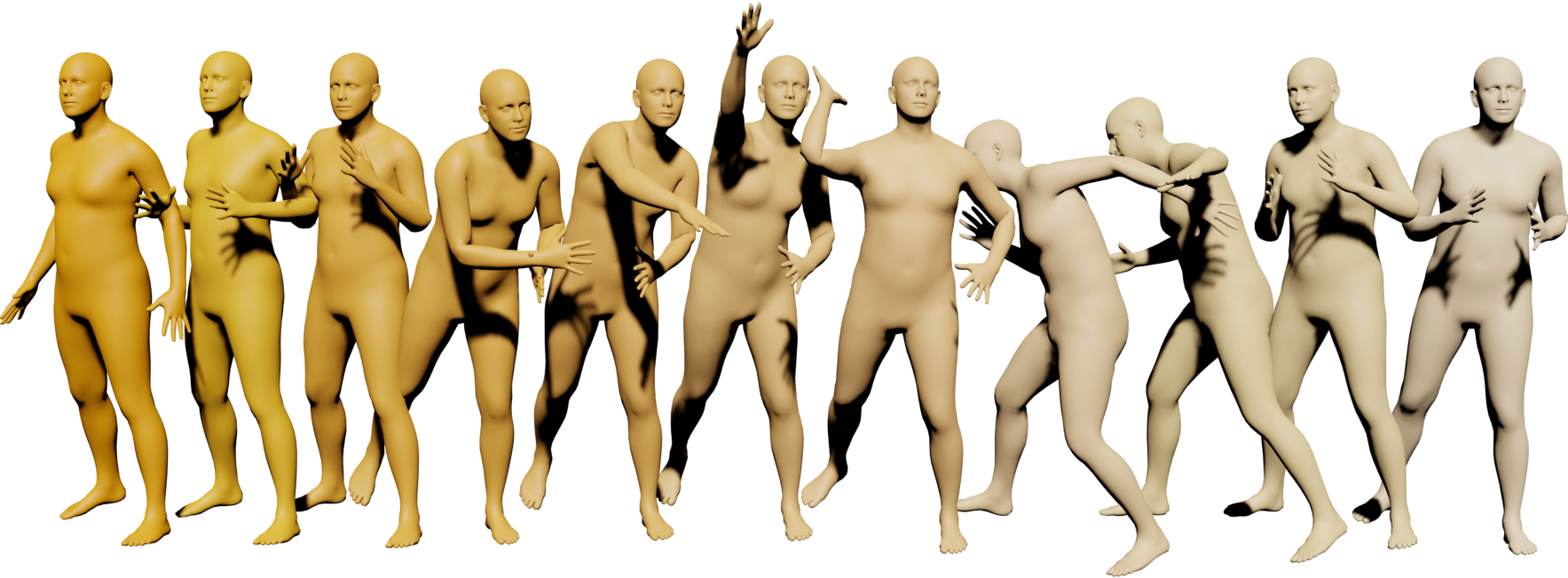}
}

\caption{Visualization on HumanML3D Dataset.}     
\label{fig: Visualization on HumanML3D Dataset} 
\end{figure}

\noindent\textbf{Comparisons on Single Motion} \quad
We compared single motion generation with SOTA methods on HumanML3D and HumanLong3D Dataset. For single motion generation, our conditional information includes the estimated motion duration and semantic information but excludes prior motion and semantic information. 
The visualization results are shown in Fig~\ref{fig: Visualization on HumanML3D Dataset}. It can be seen that AMD is capable of generating corresponding motion in response to text prompts containing a single motion while achieving smooth transitions.

\noindent\textbf{Comparisons on Music-to-Dance} \quad
We conducted comparative experiments with SOTA methods, including DanceNet~\cite{zhuang2022music2dance}, Dancerevolution~\cite{huang2020dance}, FACT~\cite{li2021ai}, and Bailando~\cite{siyao2022bailando}, on the public dataset AIST++. We employed the same data partitioning strategy as the aforementioned prior works, and we converted the data in AIST++ into HumanML3D format. 
The quantitative results are presented in Table~\ref{tab: motion synthesis evaluation on aist++}. As can be observed from the table, our method is second only to Bailando in various performance metrics, particularly in the measurement of music and dance consistency indicators BAS. Bailando employs a customized reinforcement learning module to enhance the BAS index. In contrast, our method does not incorporate any enhancements yet still achieves comparable performance to Bailando. These results demonstrate that our method not only excels in the text-to-motion task but also exhibits strong generalization capabilities to other human motion generation tasks.
\begin{table}[htb]
    \centering
    \scriptsize
    \begin{tabular}{cccccc}
    \toprule
    & \multicolumn{2}{c}{Motion Quality} & \multicolumn{2}{c}{Motion Diversity} & \\
     \cmidrule(lr){2-5}
    Method & $\operatorname{FID}_k\downarrow$ & $\operatorname{FID}_g^{\dagger}\downarrow$ &$\operatorname{Div}_k\rightarrow$ & $\operatorname{Div}_g^{\dagger}\rightarrow$ & BAS $\uparrow$ \\
    \midrule
    DanceNet & 69.18 &25.59 & 2.86 &2.85 &0.1430\\
    DanceRevolution & 73.42 &25.92 &3.52 &4.87 &0.1950\\
    FACT &35.35 &22.11 &5.94 &6.18 &0.2209 \\
    Bailando &28.16 &9.62 &7.83 &6.34 &0.2332 \\
    \midrule
    Ours  &30.28 &16.11 &6.75 &6.29 &0.2302 \\
    \bottomrule
    \end{tabular}
    \caption{Music-to-dance evaluation on AIST++ Dataset}
    \label{tab: motion synthesis evaluation on aist++}
\end{table}
\section{Conclusion}
In this paper, we present the HumanLong3D - the first dataset that pairs complex motions with long textual descriptions to address the scarcity of such data. Given the suboptimal performance of current motion generation methods on long text descriptions, we introduce a novel network architecture AMD, which combines autoregressive and diffusion models to effectively capture the information contained in long texts. Furthermore, we extend our approach to incorporate audio conditional input and construct a large-scale music-dance dataset - HumanMusic can serve as a benchmark in the field of music-to-dance.

\appendix
\bibliography{aaai24}

\end{document}